\documentclass[aps,prb,twocolumn,superscriptaddress]{revtex4}
\usepackage{epsf}
\usepackage{amsfonts}
\usepackage[fleqn]{amsmath}
\usepackage{amssymb,yfonts,mathrsfs,bbm}
\usepackage{graphicx}
\usepackage{color}

\begin{document}
\title{Collective magnetization dynamics in ferromagnetic (Ga,Mn)As mediated by photo-excited carriers}
\author{Hang Li}
\affiliation{State Key Laboratory for Superlattices and Microstructures,
Institute of Semiconductors, Chinese Academy of Sciences, Beijing 100083, China}

\author{Xinyu Liu}

\author{Ying-Yuan Zhou}
\altaffiliation[Current address: ]{Shanghai Research Center of Engineering and Technology for Solid-State Lighting, Shanhai, China}

\author{Jacek K. Furdyna}
\affiliation{Department of Physics, University of Notre Dame, Notre Dame, Indiana 46556, USA}

\author{Xinhui Zhang}
\email{xinhuiz@semi.ac.cn}
\affiliation{State Key Laboratory for Superlattices and Microstructures,
Institute of Semiconductors, Chinese Academy of Sciences, Beijing 100083, China}

\begin{abstract}
We present a study of photo-excited magnetization dynamics in ferromagnetic (Ga,Mn)As films observed by time-resolved magneto-optical measurements. The magnetization precession triggered by linearly polarized optical pulses in the absence of an external field shows a strong dependence on photon frequency when the photo-excitation energy approaches the band-edge of (Ga,Mn)As. This can be understood in terms of magnetic anisotropy modulation by both laser heating of the sample and by hole-induced non-thermal paths. Our findings provide a means for identifying the transition of laser-triggered magnetization dynamics from thermal to non-thermal mechanisms, a result that is of importance for ultrafast optical spin manipulation in ferromagnetic materials via non-thermal paths.
\end{abstract}

\maketitle
\section{\label{sec:1}Introduction}
Ultrafast manipulation of collective spin excitations in ferromagnetic materials has drawn considerable attention both for its relevance to the fundamental physics of correlated spins in non-equilibrium situations, and for its potential for spintronic information processing.\cite{1,2} The ferromagnetic semiconductor (Ga,Mn)As has been extensively investigated in this connection, since its magnetic functionality can be mediated by electrical or optical control of itinerant holes.\cite{3,4} The interest in ultrafast manipulation of magnetization in this material has in turn triggered intense research on time-resolved laser excitation of coherent magnetization precession.\cite{5,6,7,8,9,10,11,12,13,14,15,16,17}

It has been shown in earlier studies that optical excitation of magnetization precession in ferromagnetic materials originates from transient modulation of magnetic anisotropy via thermal effects (i.e., laser heating), which typically requires optical excitation energy densities of up to 1 mJ/cm$^2$.\cite{18,19,20,21} However, as previously reported for the case of (Ga,Mn)As films, excitation energy densities in the $\mu$J/cm$^2$ range were shown to be adequate for triggering coherent precession of magnetization in this material.\cite{5,6,7,8,9,10,11,12,13,14,15,16,17} One should note here that magnetic anisotropy modulation via photo-induced heating is a slow process, not really suitable for ultrafast optical manipulation of magnetization in ferromagnetic materials. Theoretical studies \cite{22} suggest, however, that non-thermal manipulation of delocalized or weakly localized holes (e.g., by changing the hole density of states by circularly-polarized laser pulses) provides an alternate method for ultrafast manipulation of magnetization in (Ga,Mn)As.

Since the influence of transient increase of hole density and of local temperature due to laser excitation take place immediately after optical pumping, both effects contribute to triggering magnetization precession in (Ga,Mn)As films. However, in earlier studies different conclusions were reported regarding the dominant effect responsible for the transient modulation of magnetic anisotropy that triggers the observed precession of magnetization.\cite{8,14,23,24} Although the non-thermal process of modulating magnetic anisotropy via photo-excited carriers has been previously suggested to be the mechanism of magnetization precession in (Ga,Mn)As,\cite{5,14,15,16} the role of such non-thermal manipulation of magnetic dynamics with time-resolved magneto-optical experiments in this material is still a controversial issue, and requires further study. In this paper we present evidence for the dependence of ultrafast magnetization dynamics on the photon energy of optical excitation observed in (Ga,Mn)As by time-resolved magneto-optical Kerr effect (TR-MOKE) experiments. A complex energy dependence of photo-excited precession frequency of magnetization was observed when the photon energy was tuned in the immediate vicinity of the (Ga,Mn)As band gap. Our results show that such modulation of magnetic anisotropy (which we ascribe to photo-excited holes) constitutes an effective mechanism for controlling the precession frequency of magnetization, thus providing experimental evidence for the possibility of non-thermal mediation of magnetic dynamics via pulsed laser excitations.

\section{\label{sec:2}Experiment}
A 97-nm thick Ga$_{0.964}$Mn$_{0.036}$As layer deposited on a GaAs (001) substrate was prepared by low-temperature molecular-beam epitaxy (LT-MBE). A piece of the sample was additionally annealed at 250$^\circ$C in N$_2$ for one hour to provide a companion sample with modified magnetic and electrical properties. The hole densities $p$ of the as-grown and annealed samples were estimated to be, respectively, $\sim$$2\times$10$^{20}$ cm$^{-3}$ and $\sim$$3\times$10$^{20}$ cm$^{-3}$, with Curie temperature $T_C$ of $\sim$58 K and $\sim$79 K as determined by superconducting quantum interference device (SQUID) measurements. The temperature dependence of the magnetization of the specimen is shown in Appendix. The TR-MOKE measurements were carried out by employing a Ti:Sapphire laser with a pulse width of 150 fs and a repetition rate of 80 MHz. The pump beam was linearly polarized, with excitation energy tuned from 1.43 eV (865 nm) to 1.81 eV (685 nm). Pump-induced changes of the magneto-optical response of the samples were measured via a time-delayed linearly polarized probe pulse. The experiments were performed in a Janis subcompact cryostat at various temperatures. No external magnetic field was applied in the experiments.

\section{\label{sec:3}Results and Analysis}
\begin{figure}[htb]
\centerline{\includegraphics[width=100mm,clip]{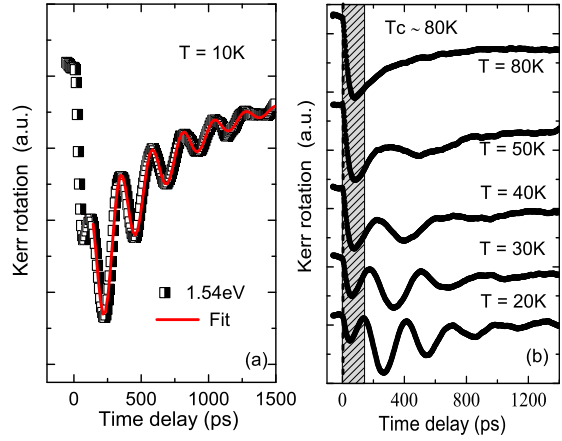}}
\caption{(Color online) (a) Temporal profile of Kerr rotation measured at 10 K for linearly- polarized pumping at 1.54 eV for the annealed (Ga,Mn)As sample. The solid line (red color) shows the best fit. (b) Time-resolved Kerr rotations excited at different ambient temperatures. The crosshatch shows that the pulse-like signal has no noticeable temperature dependence, even at temperatures above $T_c$.} \label{Fig1}
\end{figure}

Temporal evolution of the TR-MOKE response measured at 10 K with an optical excitation energy of 1.54 eV is shown in Fig.\ref{Fig1}(a), showing an initial pulse-like signal followed by exponentially damped oscillations. The initial pulse-like signal shows no dependence on temperature, persisting even to above Curie temperature, as displayed in Fig.\ref{Fig1}(b). This temperature dependence, along with its time scale in the range of tens of picoseconds, suggests that the pulse-like signal is related to the non-equilibrium electron-hole pairs in the GaAs substrate,\cite{10,25} rather than arising from ultrafast demagnetization, which is characterized by a sub-picosecond time scale.\cite{12,14}

We now focus our discussion on the oscillatory part of Fig.\ref{Fig1}(a), which represents the uniform precession of magnetization in the (Ga,Mn)As film.\cite{10} The dynamic oscillatory signal can be fitted well by an exponentially damped sine function superimposed on a pulse-like function,\cite{17}
\begin{equation}
\label{eq1}
\theta_k=a+be^{-t/t_0}+Ae^{-t/\tau_D}sin(\omega t+\phi),
\end{equation}
\noindent
where $A$, $\tau_D$, $\omega$ and $\phi$ represent, respectively, the amplitude of the oscillation, magnetization relaxation time, oscillation frequency, and the phase of the magnetization precession; and $a$ is the background offset; and $b$ and $t_0$ are the amplitude and the damping time of the pulse-like background in the slow recovery process, respectively.

\begin{figure}[htb]
\centerline{\includegraphics[width=100mm,clip]{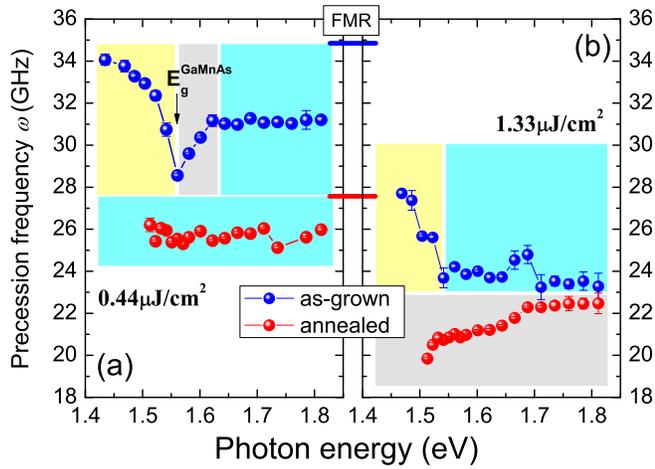}}
\caption{(Color online) Extracted precession frequencies as a function of excitation photon energy measured at 10 K with optical pumping by linearly polarized light. No external field is applied. (a) Dependence of precession frequency on photo-excitation energy for the as-grown sample (upper panel) and the annealed sample (lower panel). The black arrow represents the band edge of (Ga,Mn)As. The optical pumping intensity is 0.44 $\mu$J/cm$^2$. (b) Dependence of precession frequencies on photo-excitation energy measured with pumping intensity of 1.33 $\mu$J/cm$^2$. The lines in the middle of figure represent the precession frequency values calculated from the FMR results for the as-grown (blue) and annealed (red) samples, respectively. The color-coded regimes correspond to different dominant mechanisms responsible for the manipulation of magnetization precession as discussed in the text: the thermal effect due to laser heating (yellow regime); the nearly constant frequency resulting from the competing
role between the thermal and non-thermal effects with high density of photo-excited holes (cyan regime); the enhanced non-thermal effect due to photo-excitated holes in (Ga, Mn)As film (grey regime).} \label{Fig2}
\end{figure}

The magnetization precession frequency obtained by fitting the TR-MOKE data measured at different photo-excitation energies and pumping power densities in the absence of an external field are shown in Fig. \ref{Fig2} for both the as-grown and the annealed (Ga,Mn)As samples. We see in Fig. \ref{Fig2}(a) that the frequency of the magnetization precession of the as-grown sample exhibits a nonmonotonic dependence when the excitation energy varies from 1.43 eV to 1.81 eV: as the excitation energy increases, the precession frequency first decreases rapidly to a minimum at 1.56 eV, then increases monotonically to about 1.60 eV, and eventually levels off. It is known that the photo-excitation can cause momentary changes in both the carrier density and the sample temperature, which can then result in transient changes of the internal magnetic fields (and thus of the magnetization) in the material. Furthermore, all these changes are expected to depend on the photon energy of the optical pulse due to the variation of the absorption coefficient, especially near the bandgap.\cite{15,24}

According to the Landau-Lifshitz-Gilbert equation, the precession frequency of the magnetization is determined by the total effective magnetic field, and thus may be a function of the photon energy, as argued above. Theoretically, the effective field includes the external magnetic field, magnetic anisotropy fields, exchange field and demagnetization field.\cite{1} A transient change of this total effective field will initialize the precession of the magnetization, and will also contribute to the precession frequency. However, the exchange field itself will not affect the precession frequency because the hole spin precesses and relaxes much faster than the Mn spin.\cite{26,27} Thus, in the absence of an external magnetic field, the value of precession frequency is mainly determined by changes in the magnetic anisotropy field induced by the optical pulse.

The dependence of the precession frequency on magnetic anisotropy fields can be obtained directly from the expression for the ferromagnetic resonance (FMR) frequency.\cite{24} We recall that for thin compressively-strained (Ga,Mn)As films such as the samples used in this paper, the magnetization lies in the plane of the sample, and at low temperatures (where the cubic anisotropy is much stronger than the uniaxial anisotropy) aligns itself with the in-plane cubic easy axes, i.e., with the $<$100$>$ crystallographic directions.\cite{28} Under these conditions the precession frequency of the magnetization can be written as:\cite{29}
\begin{equation}
\label{eq2}
(\frac{\omega}{\gamma})^2=(H+H_{4\parallel})(H+4\pi M_{eff}+H_{4\parallel}+\frac{H_{2\parallel}}{2}),\
\end{equation}
\noindent
where $\gamma$ is the gyromagnetic ratio ($\gamma$ = 1.7588 Hz/Oe for $g$-factor = 2.0023), $H$ is the external magnetic field, $H_{4\parallel}$ and $H_{2\parallel}$ are the cubic and uniaxial anisotropy fields, respectively, and $4\pi M_{eff}$ is the effective perpendicular uniaxial anisotropy field, $4\pi M_{eff}=4\pi M-H_{2\perp}$, where $H_{2\perp}$ is the perpendicular uniaxial anisotropy field. In the absence of an external magnetic field, the above equation can then be simplified to:
\begin{equation}
\label{eq3}
(\frac{\omega}{\gamma})^2=H_{4\parallel}(4\pi M_{eff}+H_{4\parallel}+\frac{H_{2\parallel}}{2}).\
\end{equation}	

\begin{figure}[htb]
\centerline{\includegraphics[width=100mm,clip]{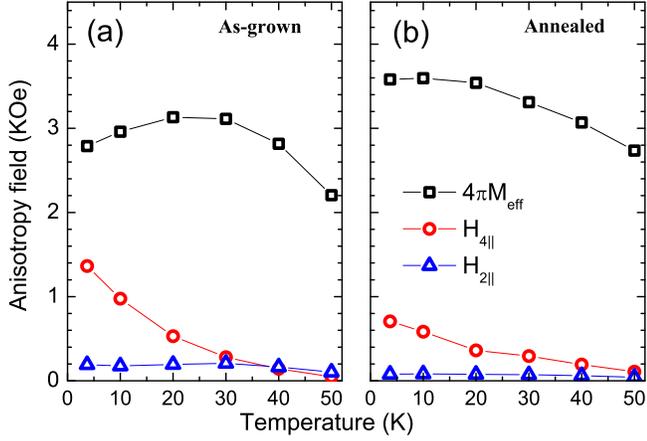}}
\caption{(Color online) Extracted magnetic anisotropy parameters for both the as-grown and annealed samples, including $4\pi M_{eff}$ and the in-plane magnetic anisotropy fields $H_{4\parallel}$ and $H_{2\parallel}$. When the temperature is above 25 K, the variation of the in-plane anisotropy fields with temperature is not obvious.} \label{Fig3}
\end{figure}

\begin{figure}[htb]
\centerline{\includegraphics[width=100mm,clip]{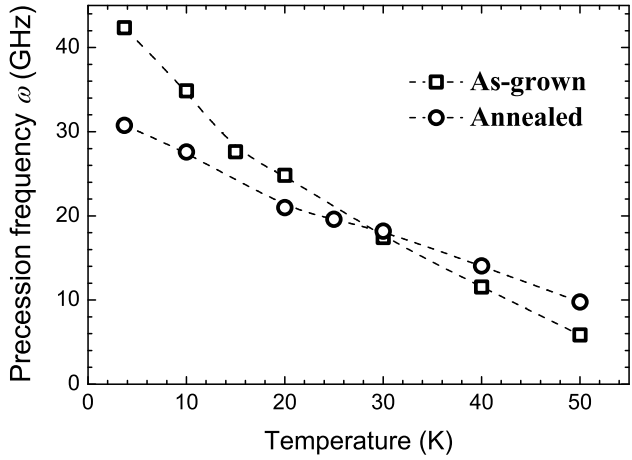}}
\caption{Calculated magnetization precession frequency as a function of temperature for both as-grown and annealed samples. The calculation shows that the increase in the sample temperature decreases the precession frequency.} \label{Fig4}
\end{figure}

In order to obtain the parameters in Eq. \ref{eq3}, we will use the results of FMR measurements carried out earlier on the same samples at a series of temperatures (see Appendix).\cite{29} The values of $4\pi M_{eff}$, $H_{4\parallel}$ and $H_{2\parallel}$ obtained by fitting the FMR results are shown in Fig. \ref{Fig3}. It is seen in the figure that, the in-plane magnetocrystalline anisotropy fields $H_{4\parallel}$ and $H_{2\parallel}$ decrease monotonically with increasing temperature, while the temperature dependence of the $4\pi M_{eff}$ shows a non-monotonic variation. The temperature dependence of the precession frequency can thus be directly obtained from the temperature dependence of the magnetic anisotropy fields. As seen in Fig. \ref{Fig4}, calculations based on Eq. \ref{eq3} and the FMR results shows that the precession frequency of the magnetization decreases monotonically with increasing temperature. This analysis clearly suggests that the precession frequency is inversely proportional to the local temperature.\cite{26,29,30,31,32} From this we conclude that, when thermal effects dominate the precession process, a transient increase of the local temperature $\Delta$$T$ induced by the absorption of an optical pulse will lead to a decrease of the precession frequency.\cite{5,6,7,8,9,10,11,12,13,14,15,16} Consistent with this expectation, in Fig. \ref{Fig2}(a) we see in that for the as-grown sample the precession frequency indeed decreases with increasing laser energy (i.e., with increase in laser-induced heating) at excitation energies below 1.56 eV, i.e., the band gap of (Ga,Mn)As,\cite{33,34} thus implying that photo-excitation-induced modulation of the precession frequency below the (Ga,Mn)As bandgap can be ascribed mainly to laser heating.

However, for excitation energies between 1.56 eV to 1.62 eV the precession frequency in the as-grown sample is clearly observed to increase with photon energy. This contrasts sharply with the behavior induced by magnetic anisotropy modulation via thermal effects just discussed. The major difference between below- and above-bandgap photo-excitations is, of course, the creation of holes, and we ascribe the observed difference in the behavior of magnetization precession to that latter effect. Indeed, it has been theoretically predicted that a change of hole density will lead to changes in magnetic anisotropy fields in (Ga,Mn)As.\cite{35,36} Furthermore, it has also been experimentally demonstrated that an increase in the hole density leads to an increase in the $4\pi M_{eff}$ parameter. Although the increase of the hole population also reduces the in-plane cubic anisotropy fields $H_{4\parallel}$ and $H_{2\parallel}$, it has been shown that the latter effect is weaker.\cite{30} This can indeed be seen in Fig. \ref{Fig3} where, for the moderate Mn concentration of $\sim$3.6$\%$ of our samples, the in-plane magnetic anisotropy fields $H_{4\parallel}$ and $H_{2\parallel}$ exhibit a decrease with the increase of hole density due to annealing, while $4\pi M_{eff}$ undergoes a noticeable increase. The striking dependence of magnetic anisotropy fields on the hole density shown in Fig. \ref{Fig3} strongly suggests that the increase of hole density due to ultrafast laser-excitation leads to a similar variation of magnetic anisotropy field.

A quantitative look at the anisotropy fields obtained from fitting the FMR data in Fig. \ref{Fig4} shows that at 10 K, for the as-grown sample the in-plane magnetic anisotropy field $H_{4\parallel}$ is two times smaller than $4\pi M_{eff}$, while for the annealed sample $H_{4\parallel}$ is six times smaller than $4\pi M_{eff}$. From this we conclude that, based on Eq. \ref{eq3}, when the change of $4\pi M_{eff}$ due to laser-induced hole density is much stronger than that of $H_{4\parallel}$, which is expected for the sample with a higher hole density,\cite{30} the variation of precession frequency is expected to be determined primarily by the trend of $4\pi M_{eff}$. One can thus readily conclude that the enhancement of the $4\pi M_{eff}$ parameter by photo-induced increase of hole density leads to an increase of precession frequency. This trend is indeed seen in Fig. \ref{Fig2}(a) for the as-grown sample at above band-edge excitations (from 1.56 eV to 1.62 eV), suggesting that the concentration of photo-excited holes plays a critical role in determining the precession frequency.

One should note, of course, that the effects of laser heating and of photo-excited carriers affect magnetization dynamics simultaneously but in opposite directions. Thus they may compensate in certain regions, resulting in a relatively constant precession frequency, as seen in Fig. \ref{Fig2}(a) for excitation energies above 1.62 eV for the as-grown sample. For completeness, we note that another possible reason for the observed leveling-off of the precession frequency at high photon excitation energies may arise as follows. It is known that the electron-hole density of states undergoes a dramatic increase between 1.56 eV and 1.62 eV near the $\Gamma$ point, but when the photo-excitation energy exceeds 1.62 eV, the electron-hole density of states quickly reaches a plateau.\cite{37} This will eventually lead to a saturation of the photo-excited carrier density, and thus to a leveling off of the precession frequency at excitation energies above 1.62 eV seen in Fig. \ref{Fig2}(a).

Figure \ref{Fig2}(b) shows the dependence of precession frequency on photo-excitation energy for a higher pump intensity. The figure clearly shows that for the as-grown (Ga,Mn)As sample a critical turning point of the precession frequency variation also occurs near the band-edge. Below the band-edge, the increased laser heating at the pump intensity of 1.33 $\mu$J/cm$^2$ causes a quicker decrease of the precession frequency, compared to the excitation at 0.44 $\mu$J/cm$^2$ seen in Fig. \ref{Fig2}(a). However, in contrast with the low-intensity results, when the excitation energy exceeds the band-edge, the precession frequency levels off at about 1.54 eV. We suggest that at this high excitation intensity the increased laser heating may be sufficient to compensate the effect of optically-pumped holes, thus resulting in a relatively flat precession frequency.

In order to further understand the dependence of magnetization precession frequency on the hole density, measurements were also carried out on the annealed sample, which has a significantly higher hole density than the as-grown specimen. Experimentally, we found that it is harder to excite the magnetization precession in the annealed sample than in the as-grown sample below the band gap. In this case one sees that at the low pumping intensity of 0.44 $\mu$J/cm$^2$ the annealing leads to a very different scenario; i.e., as shown in Fig. \ref{Fig2}(a), the precession frequency remains basically unchanged throughout the entire photon energy range used in this study. From this we conclude that in this case the effects of $H_{4\parallel}$ and $4\pi M_{eff}$ due to the increased hole concentration compensate each other. At the higher pump intensity of 1.33 $\mu$J/cm$^2$, however, the precession frequency in the annealed sample shows a continuous increase with excitation energy, as shown in Fig. 2(b). For the excitation of pumping intensity of 1.33 $\mu$J/cm$^2$, since the laser-heating-induced $\Delta$$T$ is now higher than that of lower-density excitation, the precession frequency $\omega$ at 1.51 eV drops to 19.8 GHz due to the dominance of thermal effects. However, the frequency now shows a continuous increase with photo-excitation energy from 1.52 eV to 1.81 eV. According to the discussion above, we suggest that in this case the enhanced value of $4\pi M_{eff}$ caused by the higher hole density, which continues to increase with increasing photo-excitation energy, is responsible for this behavior, thus revealing the importance of non-thermal mechanism in the annealed sample.

\begin{figure}[htb]
\centerline{\includegraphics[width=100mm,clip]{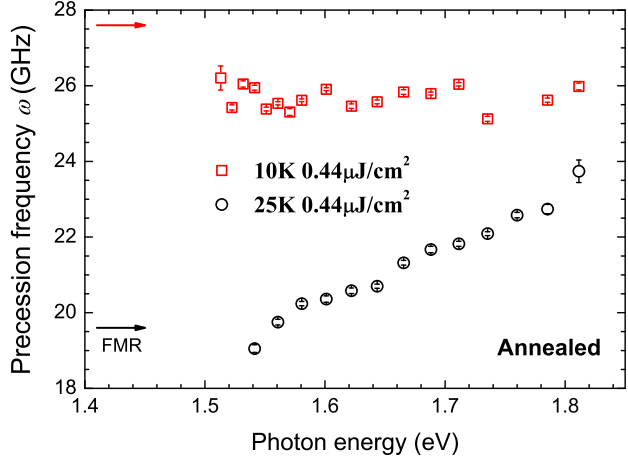}}
\caption{(Color online) Precession frequency triggered by the laser pulse as a function of photo-excitation energy measured at two temperatures at the 0.44 $\mu$J/cm$^2$ pump intensity for the annealed (Ga,Mn)As sample. The arrows represent the values calculated from the FMR results for 10 K (red) and 25 K (black), respectively.} \label{Fig5}
\end{figure}

In order to further illustrate the behavior of non-thermal effects on magnetization precession, in Fig. \ref{Fig5} we compare the photo-excitation energy dependence of the precession frequency measured at two different temperatures for the annealed sample. At 25 K the precession frequency has shown strong proportional dependence on the excitation energy with lower photo-excitation intensity of 0.44 $\mu$J/cm$^2$. Above $T$ = 25 K, since the temperature dependence of the in-plane anisotropy fields $H_{4\parallel}$ becomes not obvious as shown in Fig. \ref{Fig3}, the influence of $4\pi M_{eff}$ is more significant in the frequency analysis. As seen in Fig. \ref{Fig5}, because of the strong enhancement of $4\pi M_{eff}$ by the increase in hole density upon photo-excitation above the band-edge, the measured frequency shows a continuous increase, which is even higher than the essential value at $T$ = 25 K calculated from the FMR result. Nevertheless, as shown in Fig. \ref{Fig2}(b), at $T$ = 10 K, the optical pumping intensity must be increased to 1.33 $\mu$J/cm$^2$ to saturate the variance of $H_{4\parallel}$, so to observe the similar trend: the precession frequency increases with an increasing excitation energy.

\begin{figure}[htb]
\centerline{\includegraphics[width=100mm,clip]{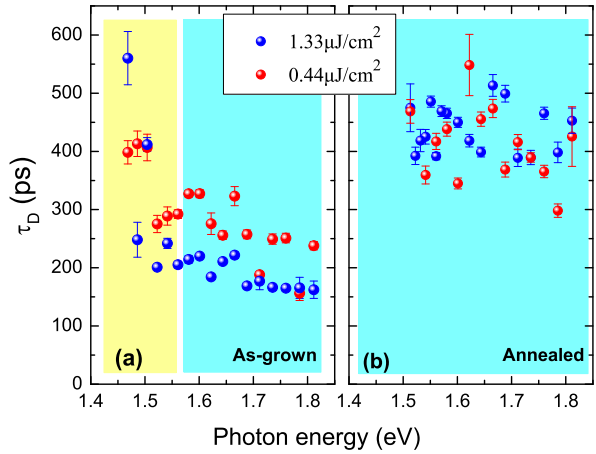}}
\caption{(Color online) (a) The magnetization relaxation time $\tau_D$ as function of photo-excitation energy measured at 10 K with linearly polarized pump pulses at 0.44 $\mu$J/cm$^2$ and 1.33 $\mu$J/cm$^2$ intensities for the as-grown sample. (b) The magnetization relaxation time $\tau_D$ as function of photo-excitation energy measured at 10 K with linearly polarized pump pulses at 0.44 $\mu$J/cm$^2$ and 1.33 $\mu$J/cm$^2$ for the annealed sample.} \label{Fig6}
\end{figure}

The impact of photo-excited carriers is also reflected in the relaxation time $\tau_D$ of the magnetization precession, which is connected to the Gilbert damping coefficient by the anisotropy fields.\cite{24} For completeness, Fig. \ref{Fig6} shows the relaxation time $\tau_D$ for both as-grown and annealed samples measured at 10 K at different optical pump intensities. As seen in Fig. \ref{Fig6}(a), below the energy gap, the magnetic relaxation time is observed to be quite strongly influenced by the photon energy for the as-grown sample. Below the band gap, various scattering processes such as hole-phonon, hole-disorder and hole-hole scatterings will be greatly reduced, since there are no spatial and temporal fluctuations created by photo-generated carriers.\cite{12,14,24} When the excitation energy is above 1.56 eV, however, the extrinsic dephasing effects due to the fluctuations created by photo-generated carriers are greatly enhanced. Thus, the relaxation time shows a clear drop, with a more obvious change at higher pumping intensity. For the annealed sample, the removal of the interstitial Mn efficiently reduces the amount of scattering source,\cite{24,38} and meanwhile, the increased background hole density can suppresses the Bir-Aronov-Pikus (BAP) spin relaxation mechanism via reducing the magnetic disorder.\cite{39,40,41} Thus, the magnetic relaxation time of the annealed sample substantially increases comparing with that of the as-grown sample. In addition, it should be mentioned that the relaxation time $\tau_D$ of the magnetization precession is also inversely proportional to the anisotropy fields.\cite{24} As shown in figure, the magnetic relaxation times for both samples exhibit negligible dependence on the excitation energy above band gap of (Ga,Mn)As. Such result indicates that the variances of $4\pi M_{eff}$ and $H_{4\parallel}$ as function of photon energy are in opposition directions and compensate each other when the photon energy is above the (Ga,Mn)As band gap, which is consistent with the results shown in Fig. \ref{Fig2}.

\section{\label{sec:4}Conclusions}
In conclusion, we have studied photo-induced magnetization dynamics in as-grown and annealed Ga$_{0.964}$Mn$_{0.036}$As films by time-resolved magneto-optical spectroscopy. The results suggest that at photo-excitation energies below the band-edge of (Ga,Mn)As the observed changes in the precession frequency arise from changes in the magnetic anisotropy fields induced through laser heating. For the regime of above-band-edge excitation, on the other hand, photo-excitation induces non-thermal effects that result from photo-excitated holes in the material. Our results reveal the competing role of these two distinct contributions in controlling the collective magnetization precession in (Ga,Mn)As, providing direct experimental evidence for the possibility of ultrafast non-thermal manipulation of magnetization dynamics in ferromagnetic (Ga,Mn)As by linearly polarized optical pulse excitation.

\begin{acknowledgments}
This work was supported by the National Basic Research Program of China (Nos.2011CB922200, 2013CB922303), and the National Natural Science Foundation of China (No.10974195). The work at Notre Dame was supported by the National Science Foundation Grant DMR14-00432.
\end{acknowledgments}

\bibliography{photocarriers}

\begin{thebibliography}{41}
\expandafter\ifx\csname natexlab\endcsname\relax\def\natexlab#1{#1}\fi
\expandafter\ifx\csname bibnamefont\endcsname\relax
  \def\bibnamefont#1{#1}\fi
\expandafter\ifx\csname bibfnamefont\endcsname\relax
  \def\bibfnamefont#1{#1}\fi
\expandafter\ifx\csname citenamefont\endcsname\relax
  \def\citenamefont#1{#1}\fi
\expandafter\ifx\csname url\endcsname\relax
  \def\url#1{\texttt{#1}}\fi
\expandafter\ifx\csname urlprefix\endcsname\relax\def\urlprefix{URL }\fi
\providecommand{\bibinfo}[2]{#2}
\providecommand{\eprint}[2][]{\url{#2}}

\bibitem[{\citenamefont{Kirilyuk et~al.}(2010)\citenamefont{Kirilyuk, Kimel,
  and Rasing}}]{1}
\bibinfo{author}{\bibfnamefont{A.}~\bibnamefont{Kirilyuk}},
  \bibinfo{author}{\bibfnamefont{A.~V.} \bibnamefont{Kimel}}, \bibnamefont{and}
  \bibinfo{author}{\bibfnamefont{T.}~\bibnamefont{Rasing}},
  \bibinfo{journal}{Rev. Mod. Phys.} \textbf{\bibinfo{volume}{82}},
  \bibinfo{pages}{2731} (\bibinfo{year}{2010}).

\bibitem[{\citenamefont{Hillebrands and Ounadjela}(2002)}]{2}
\bibinfo{author}{\bibfnamefont{B.}~\bibnamefont{Hillebrands}} \bibnamefont{and}
  \bibinfo{author}{\bibfnamefont{K.}~\bibnamefont{Ounadjela}},
  \emph{\bibinfo{title}{Spin Dynamics in Confined Magnetic Structures I}}
  (\bibinfo{publisher}{Springer Verlag, Berlin Heidelberg New York},
  \bibinfo{year}{2002}).

\bibitem[{\citenamefont{Dietl}(2010)}]{3}
\bibinfo{author}{\bibfnamefont{T.}~\bibnamefont{Dietl}}, \bibinfo{journal}{Nat.
  Mater.} \textbf{\bibinfo{volume}{9}}, \bibinfo{pages}{965}
  (\bibinfo{year}{2010}).

\bibitem[{\citenamefont{Jungwirth et~al.}(2006)\citenamefont{Jungwirth, Sinova,
  Ma\v{s}ek, Ku\v{c}era, and MacDonald}}]{4}
\bibinfo{author}{\bibfnamefont{T.}~\bibnamefont{Jungwirth}},
  \bibinfo{author}{\bibfnamefont{J.}~\bibnamefont{Sinova}},
  \bibinfo{author}{\bibfnamefont{J.}~\bibnamefont{Ma\v{s}ek}},
  \bibinfo{author}{\bibfnamefont{J.}~\bibnamefont{Ku\v{c}era}},
  \bibnamefont{and} \bibinfo{author}{\bibfnamefont{A.~H.}
  \bibnamefont{MacDonald}}, \bibinfo{journal}{Rev. Mod. Phys.}
  \textbf{\bibinfo{volume}{78}}, \bibinfo{pages}{809} (\bibinfo{year}{2006}).

\bibitem[{\citenamefont{Oiwa et~al.}(2005)\citenamefont{Oiwa, Takechi, and
  Munekata}}]{5}
\bibinfo{author}{\bibfnamefont{A.}~\bibnamefont{Oiwa}},
  \bibinfo{author}{\bibfnamefont{H.}~\bibnamefont{Takechi}}, \bibnamefont{and}
  \bibinfo{author}{\bibfnamefont{H.}~\bibnamefont{Munekata}},
  \bibinfo{journal}{J. Supercond.} \textbf{\bibinfo{volume}{18}},
  \bibinfo{pages}{9} (\bibinfo{year}{2005}).

\bibitem[{\citenamefont{Wang et~al.}(2007{\natexlab{a}})\citenamefont{Wang,
  Ren, Liu, Furdyna, Grimsditch, and Merlin}}]{6}
\bibinfo{author}{\bibfnamefont{D.~M.} \bibnamefont{Wang}},
  \bibinfo{author}{\bibfnamefont{Y.~H.} \bibnamefont{Ren}},
  \bibinfo{author}{\bibfnamefont{X.}~\bibnamefont{Liu}},
  \bibinfo{author}{\bibfnamefont{J.~K.} \bibnamefont{Furdyna}},
  \bibinfo{author}{\bibfnamefont{M.}~\bibnamefont{Grimsditch}},
  \bibnamefont{and} \bibinfo{author}{\bibfnamefont{R.}~\bibnamefont{Merlin}},
  \bibinfo{journal}{Phys. Rev. B} \textbf{\bibinfo{volume}{75}},
  \bibinfo{pages}{233308} (\bibinfo{year}{2007}{\natexlab{a}}).

\bibitem[{\citenamefont{Takechi et~al.}(2006)\citenamefont{Takechi, Oiwa,
  Nomura, Kondo, and Munekata}}]{7}
\bibinfo{author}{\bibfnamefont{H.}~\bibnamefont{Takechi}},
  \bibinfo{author}{\bibfnamefont{A.}~\bibnamefont{Oiwa}},
  \bibinfo{author}{\bibfnamefont{K.}~\bibnamefont{Nomura}},
  \bibinfo{author}{\bibfnamefont{T.}~\bibnamefont{Kondo}}, \bibnamefont{and}
  \bibinfo{author}{\bibfnamefont{H.}~\bibnamefont{Munekata}},
  \bibinfo{journal}{physica status solidi (c)} \textbf{\bibinfo{volume}{3}},
  \bibinfo{pages}{4267} (\bibinfo{year}{2006}).

\bibitem[{\citenamefont{Qi et~al.}(2007)\citenamefont{Qi, Xu, Tolk, Liu,
  Furdyna, and Perakis}}]{8}
\bibinfo{author}{\bibfnamefont{J.}~\bibnamefont{Qi}},
  \bibinfo{author}{\bibfnamefont{Y.}~\bibnamefont{Xu}},
  \bibinfo{author}{\bibfnamefont{N.~H.} \bibnamefont{Tolk}},
  \bibinfo{author}{\bibfnamefont{X.}~\bibnamefont{Liu}},
  \bibinfo{author}{\bibfnamefont{J.~K.} \bibnamefont{Furdyna}},
  \bibnamefont{and} \bibinfo{author}{\bibfnamefont{I.~E.}
  \bibnamefont{Perakis}}, \bibinfo{journal}{Appl. Phys. Lett.}
  \textbf{\bibinfo{volume}{91}}, \bibinfo{pages}{112506}
  (\bibinfo{year}{2007}).

\bibitem[{\citenamefont{Yastrubchak et~al.}(2011)\citenamefont{Yastrubchak,
  \.{Z}uk, Krzy\.{z}anowska, Domagala, Andrearczyk, Sadowski, and
  Wosinski}}]{9}
\bibinfo{author}{\bibfnamefont{O.}~\bibnamefont{Yastrubchak}},
  \bibinfo{author}{\bibfnamefont{J.}~\bibnamefont{\.{Z}uk}},
  \bibinfo{author}{\bibfnamefont{H.}~\bibnamefont{Krzy\.{z}anowska}},
  \bibinfo{author}{\bibfnamefont{J.~Z.} \bibnamefont{Domagala}},
  \bibinfo{author}{\bibfnamefont{T.}~\bibnamefont{Andrearczyk}},
  \bibinfo{author}{\bibfnamefont{J.}~\bibnamefont{Sadowski}}, \bibnamefont{and}
  \bibinfo{author}{\bibfnamefont{T.}~\bibnamefont{Wosinski}},
  \bibinfo{journal}{Phys. Rev. B} \textbf{\bibinfo{volume}{83}},
  \bibinfo{pages}{245201} (\bibinfo{year}{2011}).

\bibitem[{\citenamefont{Rozkotov\'{a}
  et~al.}(2008{\natexlab{a}})\citenamefont{Rozkotov\'{a}, N\v{e}mec,
  Horodysk\'{a}, Sprinzl, Troj\'{a}nek, Mal\'{y}, Nov\'{a}k, Olejn\'{i}k, Cukr,
  and Jungwirth}}]{10}
\bibinfo{author}{\bibfnamefont{E.}~\bibnamefont{Rozkotov\'{a}}},
  \bibinfo{author}{\bibfnamefont{P.}~\bibnamefont{N\v{e}mec}},
  \bibinfo{author}{\bibfnamefont{P.}~\bibnamefont{Horodysk\'{a}}},
  \bibinfo{author}{\bibfnamefont{D.}~\bibnamefont{Sprinzl}},
  \bibinfo{author}{\bibfnamefont{F.}~\bibnamefont{Troj\'{a}nek}},
  \bibinfo{author}{\bibfnamefont{P.}~\bibnamefont{Mal\'{y}}},
  \bibinfo{author}{\bibfnamefont{V.}~\bibnamefont{Nov\'{a}k}},
  \bibinfo{author}{\bibfnamefont{K.}~\bibnamefont{Olejn\'{i}k}},
  \bibinfo{author}{\bibfnamefont{M.}~\bibnamefont{Cukr}}, \bibnamefont{and}
  \bibinfo{author}{\bibfnamefont{T.}~\bibnamefont{Jungwirth}},
  \bibinfo{journal}{Appl. Phys. Lett.} \textbf{\bibinfo{volume}{92}},
  \bibinfo{pages}{122507} (\bibinfo{year}{2008}{\natexlab{a}}).

\bibitem[{\citenamefont{Rozkotov\'{a}
  et~al.}(2008{\natexlab{b}})\citenamefont{Rozkotov\'{a}, N\v{e}mec,
  Tesa\v{r}ov\'{a}, Mal\'{y}, Nov\'{a}k, Olejn\'{i}k, Cukr, and
  Jungwirth}}]{11}
\bibinfo{author}{\bibfnamefont{E.}~\bibnamefont{Rozkotov\'{a}}},
  \bibinfo{author}{\bibfnamefont{P.}~\bibnamefont{N\v{e}mec}},
  \bibinfo{author}{\bibfnamefont{N.}~\bibnamefont{Tesa\v{r}ov\'{a}}},
  \bibinfo{author}{\bibfnamefont{P.}~\bibnamefont{Mal\'{y}}},
  \bibinfo{author}{\bibfnamefont{V.}~\bibnamefont{Nov\'{a}k}},
  \bibinfo{author}{\bibfnamefont{K.}~\bibnamefont{Olejn\'{i}k}},
  \bibinfo{author}{\bibfnamefont{M.}~\bibnamefont{Cukr}}, \bibnamefont{and}
  \bibinfo{author}{\bibfnamefont{T.}~\bibnamefont{Jungwirth}},
  \bibinfo{journal}{Appl. Phys. Lett.} \textbf{\bibinfo{volume}{93}},
  \bibinfo{pages}{232505} (\bibinfo{year}{2008}{\natexlab{b}}).

\bibitem[{\citenamefont{Wang et~al.}(2006)\citenamefont{Wang, Sun, Hashimoto,
  Kono, Khodaparast, Cywi\'{n}ski, Sham, Sanders, Stanton, and Munekata}}]{12}
\bibinfo{author}{\bibfnamefont{J.}~\bibnamefont{Wang}},
  \bibinfo{author}{\bibfnamefont{C.}~\bibnamefont{Sun}},
  \bibinfo{author}{\bibfnamefont{Y.}~\bibnamefont{Hashimoto}},
  \bibinfo{author}{\bibfnamefont{J.}~\bibnamefont{Kono}},
  \bibinfo{author}{\bibfnamefont{G.~A.} \bibnamefont{Khodaparast}},
  \bibinfo{author}{\bibfnamefont{L.}~\bibnamefont{Cywi\'{n}ski}},
  \bibinfo{author}{\bibfnamefont{L.~J.} \bibnamefont{Sham}},
  \bibinfo{author}{\bibfnamefont{G.~D.} \bibnamefont{Sanders}},
  \bibinfo{author}{\bibfnamefont{C.~J.} \bibnamefont{Stanton}},
  \bibnamefont{and} \bibinfo{author}{\bibfnamefont{H.}~\bibnamefont{Munekata}},
  \bibinfo{journal}{J. Phys.: Condens. Matter} \textbf{\bibinfo{volume}{18}},
  \bibinfo{pages}{R501} (\bibinfo{year}{2006}).

\bibitem[{\citenamefont{Hashimoto and Munekata}(2008)}]{13}
\bibinfo{author}{\bibfnamefont{Y.}~\bibnamefont{Hashimoto}} \bibnamefont{and}
  \bibinfo{author}{\bibfnamefont{H.}~\bibnamefont{Munekata}},
  \bibinfo{journal}{Appl. Phys. Lett.} \textbf{\bibinfo{volume}{93}},
  \bibinfo{pages}{202506} (\bibinfo{year}{2008}).

\bibitem[{\citenamefont{Wang et~al.}(2007{\natexlab{b}})\citenamefont{Wang,
  Cotoros, Dani, Liu, Furdyna, and Chemla}}]{14}
\bibinfo{author}{\bibfnamefont{J.}~\bibnamefont{Wang}},
  \bibinfo{author}{\bibfnamefont{I.}~\bibnamefont{Cotoros}},
  \bibinfo{author}{\bibfnamefont{K.~M.} \bibnamefont{Dani}},
  \bibinfo{author}{\bibfnamefont{X.}~\bibnamefont{Liu}},
  \bibinfo{author}{\bibfnamefont{J.~K.} \bibnamefont{Furdyna}},
  \bibnamefont{and} \bibinfo{author}{\bibfnamefont{D.~S.}
  \bibnamefont{Chemla}}, \bibinfo{journal}{Phys. Rev. Lett.}
  \textbf{\bibinfo{volume}{98}}, \bibinfo{pages}{217401}
  (\bibinfo{year}{2007}{\natexlab{b}}).

\bibitem[{\citenamefont{Hashimoto et~al.}(2008)\citenamefont{Hashimoto,
  Kobayashi, and Munekata}}]{15}
\bibinfo{author}{\bibfnamefont{Y.}~\bibnamefont{Hashimoto}},
  \bibinfo{author}{\bibfnamefont{S.}~\bibnamefont{Kobayashi}},
  \bibnamefont{and} \bibinfo{author}{\bibfnamefont{H.}~\bibnamefont{Munekata}},
  \bibinfo{journal}{Phys. Rev. Lett.} \textbf{\bibinfo{volume}{100}},
  \bibinfo{pages}{067202} (\bibinfo{year}{2008}).

\bibitem[{\citenamefont{Kobayashi et~al.}(2010)\citenamefont{Kobayashi, Suda,
  Aoyama, Nakahara, and Munekata}}]{16}
\bibinfo{author}{\bibfnamefont{S.}~\bibnamefont{Kobayashi}},
  \bibinfo{author}{\bibfnamefont{K.}~\bibnamefont{Suda}},
  \bibinfo{author}{\bibfnamefont{J.}~\bibnamefont{Aoyama}},
  \bibinfo{author}{\bibfnamefont{D.}~\bibnamefont{Nakahara}}, \bibnamefont{and}
  \bibinfo{author}{\bibfnamefont{H.}~\bibnamefont{Munekata}},
  \bibinfo{journal}{IEEE Trans. Magn.} \textbf{\bibinfo{volume}{46}},
  \bibinfo{pages}{2470} (\bibinfo{year}{2010}).

\bibitem[{\citenamefont{N\v{e}mec et~al.}(2012)\citenamefont{N\v{e}mec,
  Rozkotov\'{a}, Tesa\v{r}ov\'{a}, Troj\'{a}nek, De~Ranieri, Olejn\'{i}k,
  Zemen, Nov\'{a}k, Cukr, Mal\'{y} et~al.}}]{17}
\bibinfo{author}{\bibfnamefont{P.}~\bibnamefont{N\v{e}mec}},
  \bibinfo{author}{\bibfnamefont{E.}~\bibnamefont{Rozkotov\'{a}}},
  \bibinfo{author}{\bibfnamefont{N.}~\bibnamefont{Tesa\v{r}ov\'{a}}},
  \bibinfo{author}{\bibfnamefont{F.}~\bibnamefont{Troj\'{a}nek}},
  \bibinfo{author}{\bibfnamefont{E.}~\bibnamefont{De~Ranieri}},
  \bibinfo{author}{\bibfnamefont{K.}~\bibnamefont{Olejn\'{i}k}},
  \bibinfo{author}{\bibfnamefont{J.}~\bibnamefont{Zemen}},
  \bibinfo{author}{\bibfnamefont{V.}~\bibnamefont{Nov\'{a}k}},
  \bibinfo{author}{\bibfnamefont{M.}~\bibnamefont{Cukr}},
  \bibinfo{author}{\bibfnamefont{P.}~\bibnamefont{Mal\'{y}}},
  \bibnamefont{et~al.}, \bibinfo{journal}{Nat. Phys.}
  \textbf{\bibinfo{volume}{8}}, \bibinfo{pages}{411} (\bibinfo{year}{2012}).

\bibitem[{\citenamefont{Beaurepaire et~al.}(1996)\citenamefont{Beaurepaire,
  Merle, Daunois, and Bigot}}]{18}
\bibinfo{author}{\bibfnamefont{E.}~\bibnamefont{Beaurepaire}},
  \bibinfo{author}{\bibfnamefont{J.-C.} \bibnamefont{Merle}},
  \bibinfo{author}{\bibfnamefont{A.}~\bibnamefont{Daunois}}, \bibnamefont{and}
  \bibinfo{author}{\bibfnamefont{J.-Y.} \bibnamefont{Bigot}},
  \bibinfo{journal}{Phys. Rev. Lett.} \textbf{\bibinfo{volume}{76}},
  \bibinfo{pages}{4250} (\bibinfo{year}{1996}).

\bibitem[{\citenamefont{Guidoni et~al.}(2002)\citenamefont{Guidoni,
  Beaurepaire, and Bigot}}]{19}
\bibinfo{author}{\bibfnamefont{L.}~\bibnamefont{Guidoni}},
  \bibinfo{author}{\bibfnamefont{E.}~\bibnamefont{Beaurepaire}},
  \bibnamefont{and} \bibinfo{author}{\bibfnamefont{J.-Y.} \bibnamefont{Bigot}},
  \bibinfo{journal}{Phys. Rev. Lett.} \textbf{\bibinfo{volume}{89}},
  \bibinfo{pages}{017401} (\bibinfo{year}{2002}).

\bibitem[{\citenamefont{Stamm et~al.}(2007)\citenamefont{Stamm, Kachel,
  Pontius, Mitzner, Quast, Holldack, Khan, Lupulescu, Aziz, Wietstruk
  et~al.}}]{20}
\bibinfo{author}{\bibfnamefont{C.}~\bibnamefont{Stamm}},
  \bibinfo{author}{\bibfnamefont{T.}~\bibnamefont{Kachel}},
  \bibinfo{author}{\bibfnamefont{N.}~\bibnamefont{Pontius}},
  \bibinfo{author}{\bibfnamefont{R.}~\bibnamefont{Mitzner}},
  \bibinfo{author}{\bibfnamefont{T.}~\bibnamefont{Quast}},
  \bibinfo{author}{\bibfnamefont{K.}~\bibnamefont{Holldack}},
  \bibinfo{author}{\bibfnamefont{S.}~\bibnamefont{Khan}},
  \bibinfo{author}{\bibfnamefont{C.}~\bibnamefont{Lupulescu}},
  \bibinfo{author}{\bibfnamefont{E.~F.} \bibnamefont{Aziz}},
  \bibinfo{author}{\bibfnamefont{M.}~\bibnamefont{Wietstruk}},
  \bibnamefont{et~al.}, \bibinfo{journal}{Nat. Mater.}
  \textbf{\bibinfo{volume}{6}}, \bibinfo{pages}{740} (\bibinfo{year}{2007}).

\bibitem[{\citenamefont{M\"{u}ller et~al.}(2009)\citenamefont{M\"{u}ller,
  Walowski, Djordjevic, Miao, Gupta, Ramos, Gehrke, Moshnyaga, Samwer,
  Schmalhorst et~al.}}]{21}
\bibinfo{author}{\bibfnamefont{G.~M.} \bibnamefont{M\"{u}ller}},
  \bibinfo{author}{\bibfnamefont{J.}~\bibnamefont{Walowski}},
  \bibinfo{author}{\bibfnamefont{M.}~\bibnamefont{Djordjevic}},
  \bibinfo{author}{\bibfnamefont{G.-X.} \bibnamefont{Miao}},
  \bibinfo{author}{\bibfnamefont{A.}~\bibnamefont{Gupta}},
  \bibinfo{author}{\bibfnamefont{A.~V.} \bibnamefont{Ramos}},
  \bibinfo{author}{\bibfnamefont{K.}~\bibnamefont{Gehrke}},
  \bibinfo{author}{\bibfnamefont{V.}~\bibnamefont{Moshnyaga}},
  \bibinfo{author}{\bibfnamefont{K.}~\bibnamefont{Samwer}},
  \bibinfo{author}{\bibfnamefont{J.}~\bibnamefont{Schmalhorst}},
  \bibnamefont{et~al.}, \bibinfo{journal}{Nat. Mater.}
  \textbf{\bibinfo{volume}{8}}, \bibinfo{pages}{56} (\bibinfo{year}{2009}).

\bibitem[{\citenamefont{Kapetanakis et~al.}(2009)\citenamefont{Kapetanakis,
  Perakis, Wickey, Piermarocchi, and Wang}}]{22}
\bibinfo{author}{\bibfnamefont{M.~D.} \bibnamefont{Kapetanakis}},
  \bibinfo{author}{\bibfnamefont{I.~E.} \bibnamefont{Perakis}},
  \bibinfo{author}{\bibfnamefont{K.~J.} \bibnamefont{Wickey}},
  \bibinfo{author}{\bibfnamefont{C.}~\bibnamefont{Piermarocchi}},
  \bibnamefont{and} \bibinfo{author}{\bibfnamefont{J.}~\bibnamefont{Wang}},
  \bibinfo{journal}{Phys. Rev. Lett.} \textbf{\bibinfo{volume}{103}},
  \bibinfo{pages}{047404} (\bibinfo{year}{2009}).

\bibitem[{\citenamefont{Hamaya et~al.}(2006{\natexlab{a}})\citenamefont{Hamaya,
  Watanabe, Taniyama, Oiwa, Kitamoto, and Yamazaki}}]{23}
\bibinfo{author}{\bibfnamefont{K.}~\bibnamefont{Hamaya}},
  \bibinfo{author}{\bibfnamefont{T.}~\bibnamefont{Watanabe}},
  \bibinfo{author}{\bibfnamefont{T.}~\bibnamefont{Taniyama}},
  \bibinfo{author}{\bibfnamefont{A.}~\bibnamefont{Oiwa}},
  \bibinfo{author}{\bibfnamefont{Y.}~\bibnamefont{Kitamoto}}, \bibnamefont{and}
  \bibinfo{author}{\bibfnamefont{Y.}~\bibnamefont{Yamazaki}},
  \bibinfo{journal}{Phys. Rev. B} \textbf{\bibinfo{volume}{74}},
  \bibinfo{pages}{045201} (\bibinfo{year}{2006}{\natexlab{a}}).

\bibitem[{\citenamefont{Qi et~al.}(2009)\citenamefont{Qi, Xu, Steigerwald, Liu,
  Furdyna, Perakis, and Tolk}}]{24}
\bibinfo{author}{\bibfnamefont{J.}~\bibnamefont{Qi}},
  \bibinfo{author}{\bibfnamefont{Y.}~\bibnamefont{Xu}},
  \bibinfo{author}{\bibfnamefont{A.}~\bibnamefont{Steigerwald}},
  \bibinfo{author}{\bibfnamefont{X.}~\bibnamefont{Liu}},
  \bibinfo{author}{\bibfnamefont{J.~K.} \bibnamefont{Furdyna}},
  \bibinfo{author}{\bibfnamefont{I.~E.} \bibnamefont{Perakis}},
  \bibnamefont{and} \bibinfo{author}{\bibfnamefont{N.~H.} \bibnamefont{Tolk}},
  \bibinfo{journal}{Phys. Rev. B} \textbf{\bibinfo{volume}{79}},
  \bibinfo{pages}{085304} (\bibinfo{year}{2009}).

\bibitem[{\citenamefont{Lochtefeld et~al.}(1996)\citenamefont{Lochtefeld,
  Melloch, Chang, and Harmon}}]{25}
\bibinfo{author}{\bibfnamefont{A.~J.} \bibnamefont{Lochtefeld}},
  \bibinfo{author}{\bibfnamefont{M.~R.} \bibnamefont{Melloch}},
  \bibinfo{author}{\bibfnamefont{J.~C.~P.} \bibnamefont{Chang}},
  \bibnamefont{and} \bibinfo{author}{\bibfnamefont{E.~S.}
  \bibnamefont{Harmon}}, \bibinfo{journal}{Appl. Phys. Lett.}
  \textbf{\bibinfo{volume}{69}}, \bibinfo{pages}{1465} (\bibinfo{year}{1996}).

\bibitem[{\citenamefont{Qi et~al.}(2008)\citenamefont{Qi, Xu, Liu, Furdyna,
  Perakis, and Tolk}}]{26}
\bibinfo{author}{\bibfnamefont{J.}~\bibnamefont{Qi}},
  \bibinfo{author}{\bibfnamefont{Y.}~\bibnamefont{Xu}},
  \bibinfo{author}{\bibfnamefont{X.}~\bibnamefont{Liu}},
  \bibinfo{author}{\bibfnamefont{J.~K.} \bibnamefont{Furdyna}},
  \bibinfo{author}{\bibfnamefont{I.~E.} \bibnamefont{Perakis}},
  \bibnamefont{and} \bibinfo{author}{\bibfnamefont{N.~H.} \bibnamefont{Tolk}},
  \bibinfo{journal}{Phys. Status Solidi (c)} \textbf{\bibinfo{volume}{5}},
  \bibinfo{pages}{2637} (\bibinfo{year}{2008}).

\bibitem[{\citenamefont{Tesa\v{r}ov\'{a}
  et~al.}(2013)\citenamefont{Tesa\v{r}ov\'{a}, check{e}mec, Rozkotov\'{a},
  Zemen, Janda, Butkovi\v{c}ov\'{a}, Troj\'{a}nek, Olejn\'{i}k, Nov\'{a}k,
  Mal\'{y} et~al.}}]{27}
\bibinfo{author}{\bibfnamefont{N.}~\bibnamefont{Tesa\v{r}ov\'{a}}},
  \bibinfo{author}{\bibfnamefont{P.}~\bibnamefont{check{e}mec}},
  \bibinfo{author}{\bibfnamefont{E.}~\bibnamefont{Rozkotov\'{a}}},
  \bibinfo{author}{\bibfnamefont{J.}~\bibnamefont{Zemen}},
  \bibinfo{author}{\bibfnamefont{T.}~\bibnamefont{Janda}},
  \bibinfo{author}{\bibfnamefont{D.}~\bibnamefont{Butkovi\v{c}ov\'{a}}},
  \bibinfo{author}{\bibfnamefont{F.}~\bibnamefont{Troj\'{a}nek}},
  \bibinfo{author}{\bibfnamefont{K.}~\bibnamefont{Olejn\'{i}k}},
  \bibinfo{author}{\bibfnamefont{V.}~\bibnamefont{Nov\'{a}k}},
  \bibinfo{author}{\bibfnamefont{P.}~\bibnamefont{Mal\'{y}}},
  \bibnamefont{et~al.}, \bibinfo{journal}{Nat. Photon.}
  \textbf{\bibinfo{volume}{7}}, \bibinfo{pages}{492} (\bibinfo{year}{2013}).

\bibitem[{\citenamefont{Welp et~al.}(2003)\citenamefont{Welp, Vlasko-Vlasov,
  Liu, Furdyna, and Wojtowicz}}]{28}
\bibinfo{author}{\bibfnamefont{U.}~\bibnamefont{Welp}},
  \bibinfo{author}{\bibfnamefont{V.~K.} \bibnamefont{Vlasko-Vlasov}},
  \bibinfo{author}{\bibfnamefont{X.}~\bibnamefont{Liu}},
  \bibinfo{author}{\bibfnamefont{J.~K.} \bibnamefont{Furdyna}},
  \bibnamefont{and}
  \bibinfo{author}{\bibfnamefont{T.}~\bibnamefont{Wojtowicz}},
  \bibinfo{journal}{Phys. Rev. Lett.} \textbf{\bibinfo{volume}{90}},
  \bibinfo{pages}{167206} (\bibinfo{year}{2003}).

\bibitem[{\citenamefont{Liu et~al.}(2003)\citenamefont{Liu, Sasaki, and
  Furdyna}}]{29}
\bibinfo{author}{\bibfnamefont{X.}~\bibnamefont{Liu}},
  \bibinfo{author}{\bibfnamefont{Y.}~\bibnamefont{Sasaki}}, \bibnamefont{and}
  \bibinfo{author}{\bibfnamefont{J.~K.} \bibnamefont{Furdyna}},
  \bibinfo{journal}{Phys. Rev. B} \textbf{\bibinfo{volume}{67}},
  \bibinfo{pages}{205204} (\bibinfo{year}{2003}).

\bibitem[{\citenamefont{Liu et~al.}(2005)\citenamefont{Liu, Lim, Dobrowolska,
  Furdyna, and Wojtowicz}}]{30}
\bibinfo{author}{\bibfnamefont{X.}~\bibnamefont{Liu}},
  \bibinfo{author}{\bibfnamefont{W.~L.} \bibnamefont{Lim}},
  \bibinfo{author}{\bibfnamefont{M.}~\bibnamefont{Dobrowolska}},
  \bibinfo{author}{\bibfnamefont{J.~K.} \bibnamefont{Furdyna}},
  \bibnamefont{and}
  \bibinfo{author}{\bibfnamefont{T.}~\bibnamefont{Wojtowicz}},
  \bibinfo{journal}{Phys. Rev. B} \textbf{\bibinfo{volume}{71}},
  \bibinfo{pages}{035307} (\bibinfo{year}{2005}).

\bibitem[{\citenamefont{Hamaya et~al.}(2006{\natexlab{b}})\citenamefont{Hamaya,
  Watanabe, Taniyama, Oiwa, Kitamoto, and Yamazaki}}]{31}
\bibinfo{author}{\bibfnamefont{K.}~\bibnamefont{Hamaya}},
  \bibinfo{author}{\bibfnamefont{T.}~\bibnamefont{Watanabe}},
  \bibinfo{author}{\bibfnamefont{T.}~\bibnamefont{Taniyama}},
  \bibinfo{author}{\bibfnamefont{A.}~\bibnamefont{Oiwa}},
  \bibinfo{author}{\bibfnamefont{Y.}~\bibnamefont{Kitamoto}}, \bibnamefont{and}
  \bibinfo{author}{\bibfnamefont{Y.}~\bibnamefont{Yamazaki}},
  \bibinfo{journal}{Phys. Rev. B} \textbf{\bibinfo{volume}{74}},
  \bibinfo{pages}{045201} (\bibinfo{year}{2006}{\natexlab{b}}).

\bibitem[{\citenamefont{Shin et~al.}(2007)\citenamefont{Shin, Chung, Lee, Liu,
  and Furdyna}}]{32}
\bibinfo{author}{\bibfnamefont{D.~Y.} \bibnamefont{Shin}},
  \bibinfo{author}{\bibfnamefont{S.~J.} \bibnamefont{Chung}},
  \bibinfo{author}{\bibfnamefont{S.}~\bibnamefont{Lee}},
  \bibinfo{author}{\bibfnamefont{X.}~\bibnamefont{Liu}}, \bibnamefont{and}
  \bibinfo{author}{\bibfnamefont{J.~K.} \bibnamefont{Furdyna}},
  \bibinfo{journal}{Phys. Rev. B} \textbf{\bibinfo{volume}{76}},
  \bibinfo{pages}{035327} (\bibinfo{year}{2007}).

\bibitem[{\citenamefont{de~Boer et~al.}(2012)\citenamefont{de~Boer, Gamouras,
  March, Nov\'ak, and Hall}}]{33}
\bibinfo{author}{\bibfnamefont{T.}~\bibnamefont{de~Boer}},
  \bibinfo{author}{\bibfnamefont{A.}~\bibnamefont{Gamouras}},
  \bibinfo{author}{\bibfnamefont{S.}~\bibnamefont{March}},
  \bibinfo{author}{\bibfnamefont{V.}~\bibnamefont{Nov\'ak}}, \bibnamefont{and}
  \bibinfo{author}{\bibfnamefont{K.~C.} \bibnamefont{Hall}},
  \bibinfo{journal}{Phys. Rev. B} \textbf{\bibinfo{volume}{85}},
  \bibinfo{pages}{033202} (\bibinfo{year}{2012}).

\bibitem[{\citenamefont{Tesa\v{r}ov\'{a}
  et~al.}(2014)\citenamefont{Tesa\v{r}ov\'{a}, Ostatnick\'{y}, Nov\'{a}k,
  Olejn\'{i}k, \v{S}ubrt, Reichlov\'{a}, Ellis, Mukherjee, Lee, Sipahi
  et~al.}}]{34}
\bibinfo{author}{\bibfnamefont{N.}~\bibnamefont{Tesa\v{r}ov\'{a}}},
  \bibinfo{author}{\bibfnamefont{T.}~\bibnamefont{Ostatnick\'{y}}},
  \bibinfo{author}{\bibfnamefont{V.}~\bibnamefont{Nov\'{a}k}},
  \bibinfo{author}{\bibfnamefont{K.}~\bibnamefont{Olejn\'{i}k}},
  \bibinfo{author}{\bibfnamefont{J.}~\bibnamefont{\v{S}ubrt}},
  \bibinfo{author}{\bibfnamefont{H.}~\bibnamefont{Reichlov\'{a}}},
  \bibinfo{author}{\bibfnamefont{C.~T.} \bibnamefont{Ellis}},
  \bibinfo{author}{\bibfnamefont{A.}~\bibnamefont{Mukherjee}},
  \bibinfo{author}{\bibfnamefont{J.}~\bibnamefont{Lee}},
  \bibinfo{author}{\bibfnamefont{G.~M.} \bibnamefont{Sipahi}},
  \bibnamefont{et~al.}, \bibinfo{journal}{Phys. Rev. B}
  \textbf{\bibinfo{volume}{89}}, \bibinfo{pages}{085203}
  (\bibinfo{year}{2014}).

\bibitem[{\citenamefont{Dietl et~al.}(2001)\citenamefont{Dietl, Ohno, and
  Matsukura}}]{35}
\bibinfo{author}{\bibfnamefont{T.}~\bibnamefont{Dietl}},
  \bibinfo{author}{\bibfnamefont{H.}~\bibnamefont{Ohno}}, \bibnamefont{and}
  \bibinfo{author}{\bibfnamefont{F.}~\bibnamefont{Matsukura}},
  \bibinfo{journal}{Phys. Rev. B} \textbf{\bibinfo{volume}{63}},
  \bibinfo{pages}{195205} (\bibinfo{year}{2001}).

\bibitem[{\citenamefont{Zemen et~al.}(2009)\citenamefont{Zemen, Ku\v{c}era,
  Olejn\'{i}k, and Jungwirth}}]{36}
\bibinfo{author}{\bibfnamefont{J.}~\bibnamefont{Zemen}},
  \bibinfo{author}{\bibfnamefont{J.}~\bibnamefont{Ku\v{c}era}},
  \bibinfo{author}{\bibfnamefont{K.}~\bibnamefont{Olejn\'{i}k}},
  \bibnamefont{and}
  \bibinfo{author}{\bibfnamefont{T.}~\bibnamefont{Jungwirth}},
  \bibinfo{journal}{Phys. Rev. B} \textbf{\bibinfo{volume}{80}},
  \bibinfo{pages}{155203} (\bibinfo{year}{2009}).

\bibitem[{\citenamefont{Kapetanakis et~al.}(2012)\citenamefont{Kapetanakis,
  Wang, and Perakis}}]{37}
\bibinfo{author}{\bibfnamefont{M.~D.} \bibnamefont{Kapetanakis}},
  \bibinfo{author}{\bibfnamefont{J.}~\bibnamefont{Wang}}, \bibnamefont{and}
  \bibinfo{author}{\bibfnamefont{I.~E.} \bibnamefont{Perakis}},
  \bibinfo{journal}{J. Opt. Soc. Am. B} \textbf{\bibinfo{volume}{29}},
  \bibinfo{pages}{A95} (\bibinfo{year}{2012}).

\bibitem[{\citenamefont{Dietl et~al.}(2008)\citenamefont{Dietl, Awschalom,
  Kaminska, and Ohno}}]{38}
\bibinfo{author}{\bibfnamefont{T.}~\bibnamefont{Dietl}},
  \bibinfo{author}{\bibfnamefont{D.~D.} \bibnamefont{Awschalom}},
  \bibinfo{author}{\bibfnamefont{M.}~\bibnamefont{Kaminska}}, \bibnamefont{and}
  \bibinfo{author}{\bibfnamefont{H.}~\bibnamefont{Ohno}},
  \emph{\bibinfo{title}{Spintronics}} (\bibinfo{publisher}{Elsevier,
  Amsterdam}, \bibinfo{year}{2008}).

\bibitem[{\citenamefont{Astakhov et~al.}(2008)\citenamefont{Astakhov, Dzhioev,
  Kavokin, Korenev, Lazarev, Tkachuk, Kusrayev, Kiessling, Ossau, and
  Molenkamp}}]{39}
\bibinfo{author}{\bibfnamefont{G.~V.} \bibnamefont{Astakhov}},
  \bibinfo{author}{\bibfnamefont{R.~I.} \bibnamefont{Dzhioev}},
  \bibinfo{author}{\bibfnamefont{K.~V.} \bibnamefont{Kavokin}},
  \bibinfo{author}{\bibfnamefont{V.~L.} \bibnamefont{Korenev}},
  \bibinfo{author}{\bibfnamefont{M.~V.} \bibnamefont{Lazarev}},
  \bibinfo{author}{\bibfnamefont{M.~N.} \bibnamefont{Tkachuk}},
  \bibinfo{author}{\bibfnamefont{Y.~G.} \bibnamefont{Kusrayev}},
  \bibinfo{author}{\bibfnamefont{T.}~\bibnamefont{Kiessling}},
  \bibinfo{author}{\bibfnamefont{W.}~\bibnamefont{Ossau}}, \bibnamefont{and}
  \bibinfo{author}{\bibfnamefont{L.~W.} \bibnamefont{Molenkamp}},
  \bibinfo{journal}{Phys. Rev. Lett.} \textbf{\bibinfo{volume}{101}},
  \bibinfo{pages}{076602} (\bibinfo{year}{2008}).

\bibitem[{\citenamefont{Bir et~al.}(1975)\citenamefont{Bir, Aronov, and
  Pikus}}]{40}
\bibinfo{author}{\bibfnamefont{G.~L.} \bibnamefont{Bir}},
  \bibinfo{author}{\bibfnamefont{A.~G.} \bibnamefont{Aronov}},
  \bibnamefont{and} \bibinfo{author}{\bibfnamefont{G.~E.} \bibnamefont{Pikus}},
  \bibinfo{journal}{Zh. Eksp. Teor. Fiz.} \textbf{\bibinfo{volume}{69}},
  \bibinfo{pages}{1382} (\bibinfo{year}{1975}).

\bibitem[{\citenamefont{Bir et~al.}(1976)\citenamefont{Bir, Aronov, and
  Pikus}}]{41}
\bibinfo{author}{\bibfnamefont{G.~L.} \bibnamefont{Bir}},
  \bibinfo{author}{\bibfnamefont{A.~G.} \bibnamefont{Aronov}},
  \bibnamefont{and} \bibinfo{author}{\bibfnamefont{G.~E.} \bibnamefont{Pikus}},
  \bibinfo{journal}{Sov. Phys. JETP} \textbf{\bibinfo{volume}{42}},
  \bibinfo{pages}{705} (\bibinfo{year}{1976}).

\end{thebibliography}

\newpage

\setcounter{figure}{0}
\renewcommand{\thefigure}{A\arabic{figure}}

\begin{appendix}

\begin{figure}[htb]

\raggedright\large\textbf{Appendix}
~\\[2cm]

\centerline{\includegraphics[width=80mm,clip]{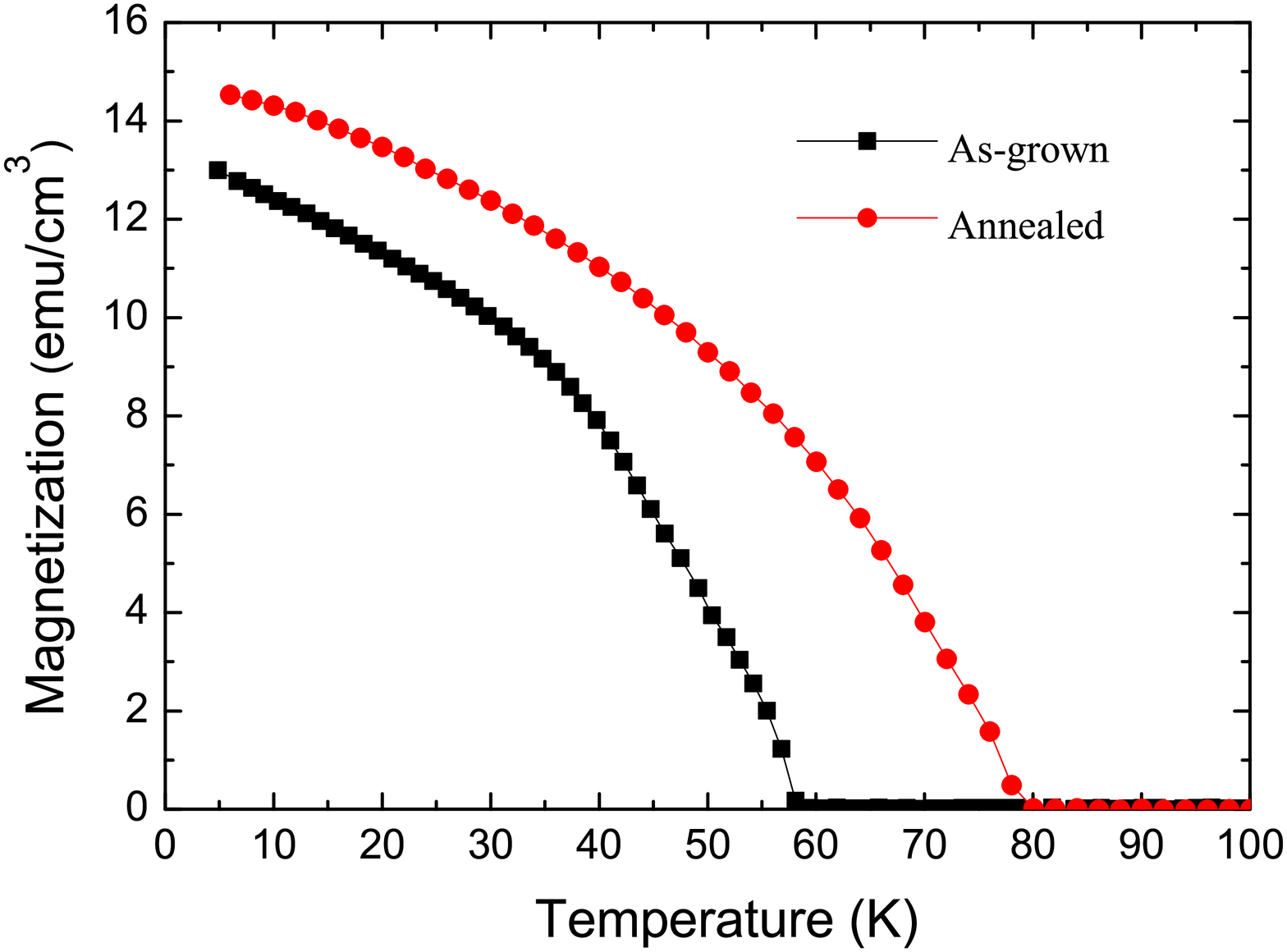}}
\caption{M-T curves for the as-grown and annealed samples. The experiments show the Curie temperatures of the samples are 58 K and 79K, respectively.} \label{sfig01}
\end{figure}

~\\[3cm]
\begin{figure}[htb]
\centerline{\includegraphics[width=80mm,clip]{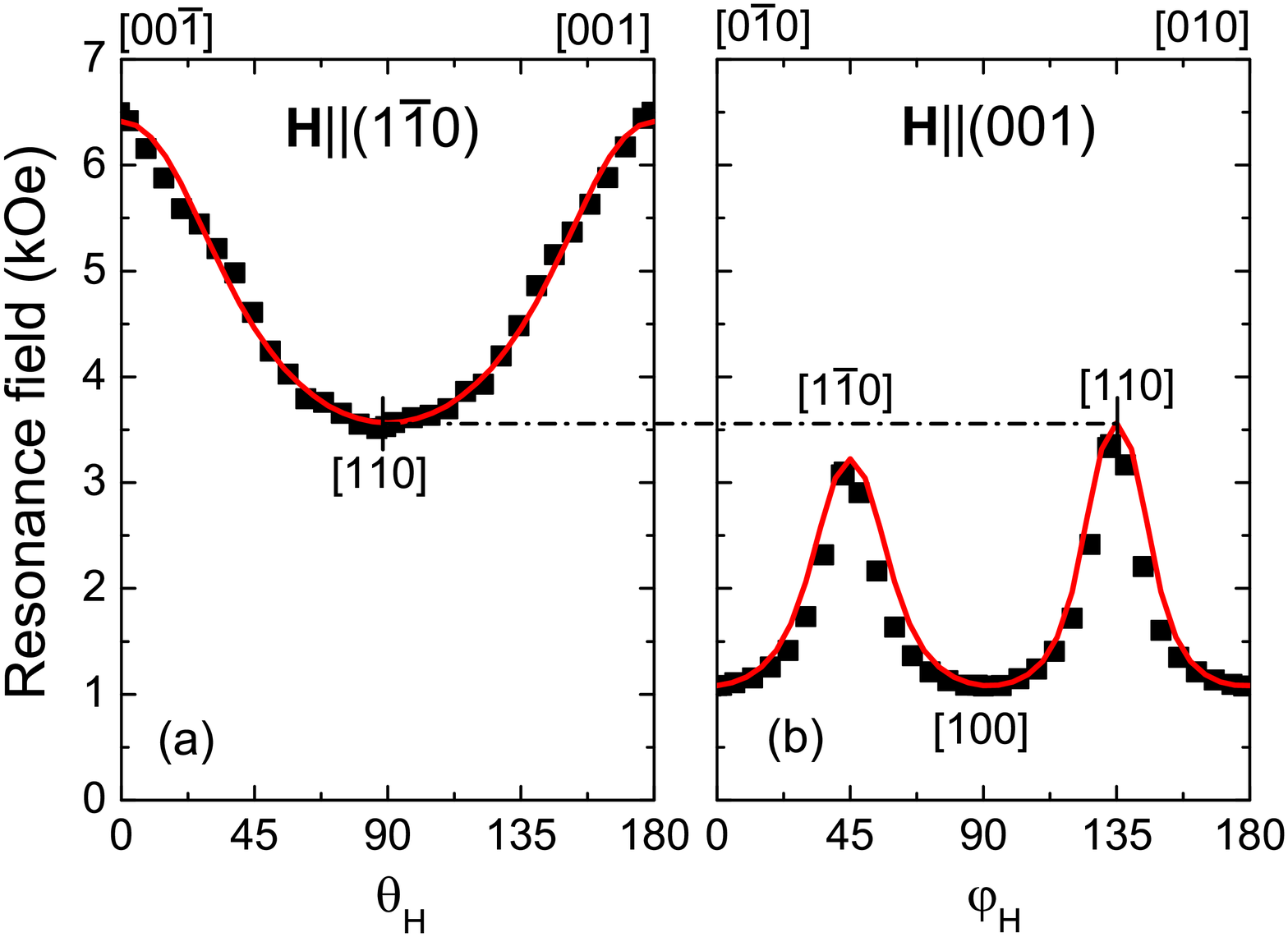}}
\caption{FMR results for the as-grown sample at $T$ = 4 K. Red solid lines represent best fit results, from which the values of anisotropy fields are extracted for the as-grown specimen.} \label{sfig02}
\end{figure}

~\\[3cm]
\begin{figure}[htb]
\centerline{\includegraphics[width=80mm,clip]{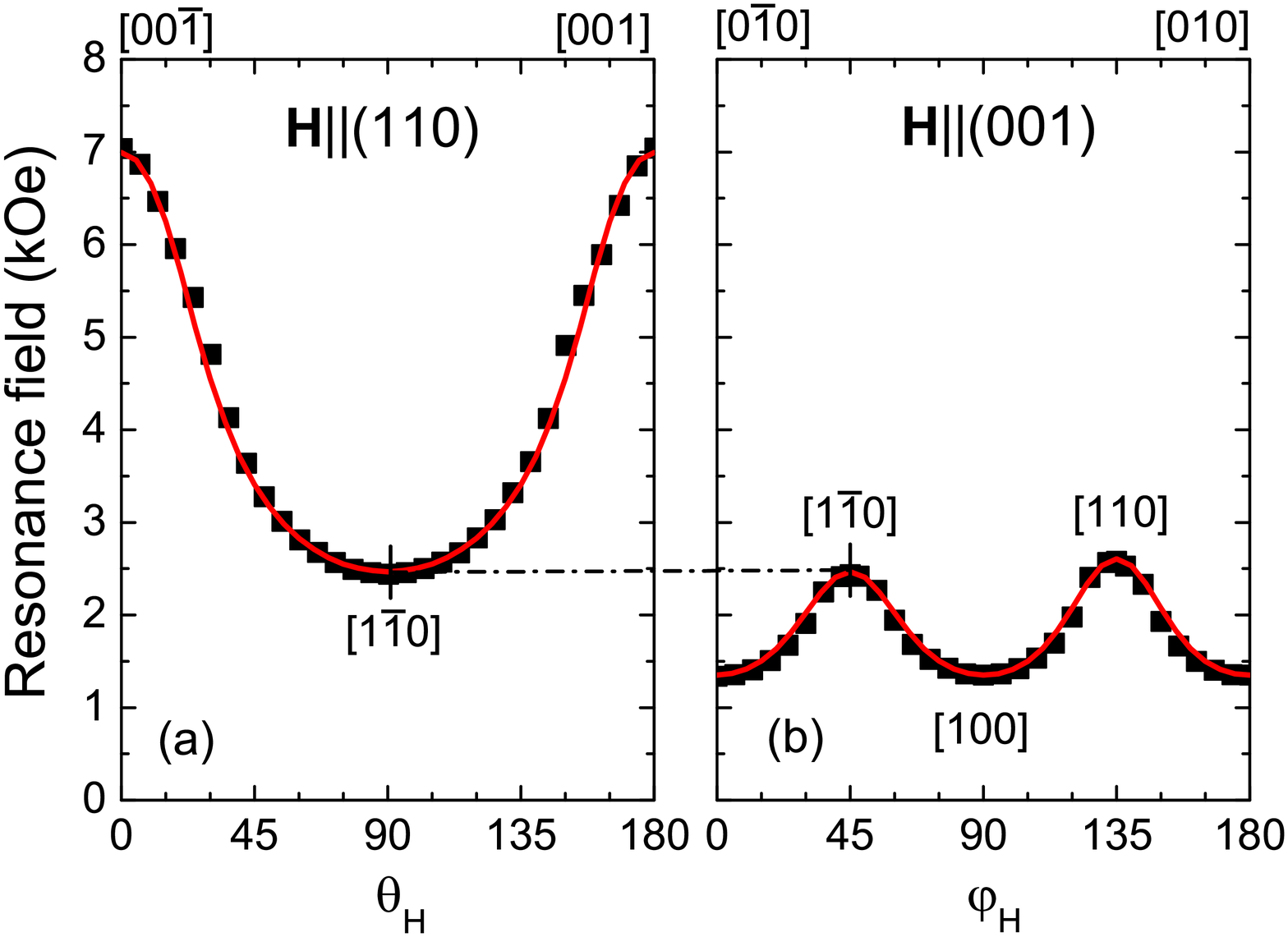}}
\caption{FMR results for the annealed sample at $T$ = 4 K. Red solid lines are the best-fit results, from which the values of anisotropy fields are extracted for the annealed specimen.} \label{sfig03}
\end{figure}

\end{appendix}
\end{document}